\documentclass{elsart}

\def\ga{\gamma}
\def\de{\delta}

\def\si{\sigma}

\def\ps{\psi}
\def\om{\omega}
\def\Ga{\Gamma}

\def\mn{{\mu\nu}}

\def\half{{\textstyle{1\over 2}}}
\def\frac#1#2{{\textstyle{{#1}\over {#2}}}}

\def\lsim{\mathrel{\rlap{\lower4pt\hbox{\hskip1pt$\sim$}}
    \raise1pt\hbox{$<$}}}
\def\gsim{\mathrel{\rlap{\lower4pt\hbox{\hskip1pt$\sim$}}
    \raise1pt\hbox{$>$}}}
\def\sqr#1#2{{\vcenter{\vbox{\hrule height.#2pt
         \hbox{\vrule width.#2pt height#1pt \kern#1pt
         \vrule width.#2pt}
         \hrule height.#2pt}}}}

\newcommand{\beq}{\begin{equation}}
\newcommand{\eeq}{\end{equation}}
\newcommand{\bea}{\begin{eqnarray}}
\newcommand{\eea}{\end{eqnarray}}

\renewenvironment{thebibliography}[1]
 { \rm
   \begin{list}{\arabic{enumi}.}
    {\usecounter{enumi} \setlength{\parsep}{0pt}
     \setlength{\itemsep}{3pt} \settowidth{\labelwidth}{#1.}
     \sloppy
    }}{\end{list}}

\usepackage{amssymb}

\begin{document}

\begin{frontmatter}



\title{Lorentz and CPT Tests in \\
Matter and Antimatter}

\author[label1]{Robert Bluhm}
\address[label1]{Department of Physics and Astronomy\\
Colby College, Waterville, ME 04901, USA}

\author{}

\address{}

\begin{abstract}
A review of recent theoretical work investigating tests of Lorentz 
and CPT symmetry in atomic and particle systems is presented.  
A variety of tests in matter and antimatter are discussed, 
including measurements of anomalous magnetic moments in Penning traps, 
comparisons of atomic-clock transitions, 
high-precision spectroscopic measurements of hydrogen and antihydrogen, 
experiments with muons, 
experiments with mesons,
and tests of Lorentz symmetry with a spin-polarized torsion pendulum.
\end{abstract}

\end{frontmatter}


\section{Introduction}

All physical interactions appear to be invariant under both
Lorentz and CPT transformations.
These symmetries are linked by the CPT theorem
\cite{cptthm},
which in essence states that local
relativistic field theories of point particles are symmetric under CPT.
A related theorem has also recently been proved
\cite{owg}.
It states that that when a field theory violates CPT it also
violates Lorentz symmetry.
Numerous experiments confirm Lorentz and CPT symmetry to
extremely high precision.
For example,
Hughes-Drever type experiments are generally considered the
best tests of Lorentz symmetry.
These experiments place very tight bounds on spatially anisotropic interactions
\cite{cctests}.

Despite the strong theoretical and experimental support for 
these symmetries,
there has been a growing interest in testing Lorentz and CPT 
symmetry in recent years
\cite{cpt01}.
This is motivated by both theoretical developments
and improved experimental capabilities.
For example,
it has been shown that string theory can
lead to violations of Lorentz and CPT symmetry
\cite{kskp}.
This is because strings are nonpointlike and have nonlocal interactions
and can therefore evade the CPT theorem.
In particular,
there are mechanisms in string theory that can induce
spontaneous breaking of Lorentz and CPT symmetry.
It has also been shown that noncommutative geometries
can arise naturally in string theory and that Lorentz violation 
is intrinsic to noncommutative field theories
\cite{ncqed}.
Lorentz violation has also been proposed as 
a breakdown of quantum mechanics in gravity
\cite{hawk82},
or as a feature of certain 
non-string approaches to quantum gravity
\cite{qg}.

Experimental signals due to effects in these kinds of 
theories are expected at the Planck scale,
$M_{\rm Pl} = \sqrt{\hbar c/G} \simeq 10^{19}$ GeV.
One approach to probing the Planck scale is to adopt Lorentz and CPT
violation as a candidate signal of new physics originating
from the Planck scale.
In this approach,
experiments search for effects that are heavily suppressed
at ordinary energies,
e.g., with suppression factors proportional to the ratio of a 
low-energy scale to the Planck scale.
Normally,
such heavily suppressed signals would be unobservable.
However,
with a unique signal such as Lorentz or CPT violation
(which cannot be mimicked in conventional physics)
it becomes possible to search for effects originating from the Planck scale.
This approach to testing Planck-scale physics has been greatly aided  
by the development the standard-model extension (SME)
\cite{ck}.
In the context of the SME,
it is possible to look for new signatures of Lorentz and CPT violation
in atomic and particle systems that might otherwise be overlooked.

Experiments in atomic systems are very well suited to this approach
because they can be sensitive to extremely low energies.
Sensitivity to frequency shifts at the level of 1 mHz or less
are routinely attained.
Expressing this as an energy shift in GeV
corresponds to a sensitivity of roughly
$4 \times 10^{-27}$ GeV.
This sensitivity is well within the range of energy one
would associate with suppression factors originating
from the Planck scale.
For example,
the ratio $m_e/M_{\rm Pl}$ multiplying the electron mass
yields an energy of approximately $2.5 \times 10^{-26}$ GeV.

Some specific examples of experiments 
that are highly sensitive to Lorentz and CPT violation 
include experiments with
electrons
\cite{dehmelt99,mittleman99,bkr9798,bk00,eotwash},
muons
\cite{muonium01,muong01,bkl00},
protons
\cite{Hmaser,bkr99},
neutrons
\cite{dualmaser},
mesons
\cite{mesonexpt,mesonth},
and photons
\cite{photons}.
These include several classic tests of Lorentz and CPT symmetry,
such as $g-2$ experiments in Penning traps
\cite{penningtests}
and atomic-clock comparisons -- the so-called 
Hughes-Drever experiments
\cite{cctests,kl99,bklr02}.

\section{Standard-Model Extension}

At low energies relative to the Planck scale,
observable effects of Lorentz and CPT violation 
are described by the Standard-Model Extension (SME)
\cite{ck}.
In full generality,
the SME allows for all coordinate-independent violations of Lorentz symmetry 
in a quantum field theory and provides a connection to the Planck scale 
through operators of nonrenormalizable dimension 
\cite{kle}.
To consider experiments in atomic physics it suffices
to restrict the SME to its QED sector and to include 
only terms that are power-counting renormalizable. 
The resulting QED extension has energy-momentum conservation,
the usual spin-statistics connection, and observer Lorentz covariance.

The modified Dirac equation in the QED extension describing a
four-component spinor field $\ps$
of mass $m$
and charge $q = -|e|$ in an electric potential $A^\mu$ is
\beq
( i \Ga^\mu D_\mu - M) \ps = 0
\quad ,
\label{dirac}
\eeq
where
$\Ga_\nu = \ga_\nu + c_\mn \ga^\mu + d_\mn \ga_5 \ga^\mu$
and
$M = m + a_\mu \ga^\mu + b_\mu \ga_5 \ga^\mu + \half H_\mn \si^\mn$.
Here,
natural units with $\hbar = c = 1$ are used,
and $i D_\mu \equiv i \partial_\mu - q A_\mu$.
The two terms involving the effective coupling constants
$a_\mu$ and $b_\mu$ violate CPT,
while the three terms involving
$H_{\mu \nu}$, $c_{\mu \nu}$, and $d_{\mu \nu}$
preserve CPT.
All five terms break Lorentz symmetry.

The recent atomic experiments that test Lorentz and CPT symmetry
express the bounds they obtain in terms of these parameters
$a_\mu$, $b_\mu$, $c_{\mu \nu}$, $d_{\mu \nu}$, and $H_{\mu \nu}$.
This provides a straightforward way of making comparisons across different
types of experiments and avoids problems that can arise
when different physical quantities
($g$ factors, charge-to-mass ratios, masses, frequencies, etc.)
are used in different experiments.
Each different particle sector in the QED extension
has a set of Lorentz-violating parameters that are independent.
The parameters of the different sectors are distinguished
using superscript labels.

\section{Tests in Atomic Systems}

Before discussing recent experiments individually,
it is useful to examine some of the more general
results that have emerged from these investigations.
One general feature that has emerged is that
the sensitivity to Lorentz and CPT violation in these experiments
stems primarily from their
ability to detect very small anomalous energy shifts.
While many of the experiments were originally designed to
measure specific quantities,
such as differences in $g$ factors or
charge-to-mass ratios of particles and antiparticles,
it is now recognized that they are most effective as
Lorentz and CPT tests when all of the energy levels in the system
are investigated for possible anomalous shifts.
Because of this,
several new signatures of Lorentz and CPT violation have been
discovered in recent years that were overlooked previously.
A second general feature is that these atomic experiments can 
be divided into two groups.
The first (traditional Lorentz tests) looks for sidereal time
variations in the energy levels of a particle or atom.
The second (traditional CPT test) looks for a difference in
the energy levels between a particle (or atom) and its
antiparticle (or antiatom).
The sensitivity to Lorentz and CPT violation in these
two classes of experiments, however, are not distinct.
Experiments traditionally viewed as CPT tests are
also sensitive to Lorentz symmetry and vice versa.
Nonetheless,
it is important to keep in mind that CPT experiments comparing
matter and antimatter are directly sensitive to CPT-violating
parameters, such as $b_\mu$,
whereas Lorentz tests are sensitive to combinations of 
CPT-preserving and CPT-violating parameters,
which we denote below using a tilde.
In this respect,
both clases of experiments should be viewed as complementary.

{\it Penning-Trap Experiments}:
The aim of the original experiments with Penning traps was to
make high-precision comparisons of the $g$ factors and charge-to-mass ratios of
particles and antiparticles confined within the trap
\cite{penningtests}.
This was obtained through measurements of the
anomaly frequency $\om_a$ and the cyclotron frequency $\om_c$.
For example,
$g-2=2\om_a/\om_c$.
The frequencies were typically measured to $\sim 10^{-9}$ for the electron,
which determines $g$ to $\sim 10^{-12}$.
In computing these ratios it was not necessary to keep
track of the times when $\om_a$ and $\om_c$ were measured.
More recently,
however,
additional signals of possible Lorentz and CPT violation
in this system have been found,
which has led to two new tests being performed.

The first was a Lorentz test involving data only for the electron
\cite{mittleman99}.
It involved looking for small sidereal time variations in the
electron anomaly frequency as the Earth turns about its axis.
The bounds in this case are given with respect to a
nonrotating coordinate system such as celestial equatorial coordinates.
The interactions involve a combination of laboratory-frame components
that couple to the electron spin.
The combination is denoted as
$\tilde b_3^e  \equiv b_3^e - m d_{30}^e - H_{12}^e$.
The bound can be expressed in terms of components $X$, $Y$, $Z$ 
in the nonrotating frame.
It is given as
$|\tilde b_J^e| \lsim 5 \times 10^{-25} {\rm GeV}$ for $J=X,Y$.

The second was a CPT test comparing particles and antiparticles.
It was a reanalysis performed by Dehmelt's group of existing
data for electrons and positrons in a Penning trap
\cite{dehmelt99}.
The idea was to search for an instantaneous difference in the
anomaly frequencies of electrons and positrons.
The original measurements of $g-2$
did not involve looking for possible instantaneous variations in $\om_a$.
Instead,
the ratio $\om_a/\om_c$ was obtained using averaged values.
The new analysis is especially relevant because it can be shown that
the CPT-violating corrections to the anomaly frequency
$\om_a$ can occur even though the $g$ factor remains unchanged.
The new bound found by Dehmelt's group can be expressed in terms of 
the parameter $b^e_3$,
which is the component of $b^e_\mu$ along the quantization
axis in the laboratory frame.
They obtained $|b^e_3| \lsim 3 \times 10^{-25}$ GeV.

{\it Clock-Comparison Experiments}:
The Hughes-Drever experiments
are classic tests of Lorentz invariance
\cite{cctests}.
These experiments look for relative changes between two atomic ``clock''
frequencies as the Earth rotates.
The ``clock'' frequencies are typically atomic hyperfine or Zeeman transitions.
Recently,
several new clock-comparison tests have been performed
or are in the planning stages.
For example,
Bear {\it et al.} have used a two-species noble-gas maser to
test for Lorentz and CPT violation in the neutron sector
\cite{dualmaser}.
They obtained a bound
$|\tilde b_J^n| \lsim 10^{-31} {\rm GeV}$ for $J=X,Y$.
It should be kept in mind that certain assumptions about the nuclear
configurations must be made to obtain these bounds.
For this reason,
they should be viewed as accurate to within one or two
orders of magnitude.
To obtain cleaner bounds it is necessary to consider
simpler atoms or to perform more precise nuclear modeling.

{\it Hydrogen/Antihydrogen Experiments}:
A recent Lorentz and CPT test in hydrogen has been performed,
and two experiments are underway at CERN to perform high-precision
tests in antihydrogen.

The experiment in hydrogen looked for sidereal time variations
in ground-state Zeeman hyperfine transitions using a hydrogen maser 
and a double-resonance technique
\cite{Hmaser}.
It yielded sharp new bounds for the electron and proton.
The bound obtained for the proton was $|\tilde b_J^p| \lsim 10^{-27}$ GeV.
Due to the simplicity of the hydrogen nucleus,
this is an extremely clean bound.

Two experiments at CERN aim to
make high-precision spectroscopic measurements of the 1S-2S
transitions in hydrogen and antihydrogen.
These are forbidden two-photon transitions with a relative linewidth
of approximately $10^{-15}$.
The magnetic field plays an important role
in the sensitivity of these transitions to Lorentz and CPT breaking.
For example,
in free hydrogen in the absence of a magnetic field,
the 1S and 2S levels shift by the same amount at leading order,
and there are no leading-order corrections to the 1S-2S transition.
However,
in a magnetic trap
there are fields that mix the different spin states in the
four hyperfine levels.
Since the Lorentz-violating couplings are spin-dependent,
there will be sensitivity at leading order
to Lorentz and CPT violation in comparisons of 1S-2S transitions.
An alternative to 1S-2S transitions is to consider 
measurements of ground-state Zeeman hyperfine transitions.
These measurments should be able to provide a clean
measurement of the CPT-violating parameter $b_\mu^p$ for the proton.
For example,
comparing Zeeman transitions in hydrogen and antihydrogen
at a field-independent point with $B \simeq 0.65$ T with
a resolution of 1 mHz would give rise to a bound on $b_3^p$
at the level of $10^{-27}$ GeV.

{\it Muon Experiments}
There are two recent experiments with muons that have sharp
sensitivity to Lorentz and CPT violation.
The first is in muonium.
It looks for sidereal time variations in the frequencies
of ground-state Zeeman hyperfine transitions
in a strong magnetic field.
A bound at a level of $| \tilde b^\mu_J| \le 2 \times 10^{-23}$ GeV
has been obtained from these measurements
\cite{muonium01}.
The second experiment is the Brookhaven $g-2$ experiments with positive muons
\cite{muong01}.
It uses relativistic muons with a ``magic'' boost 
parameter $\de = 29.3$.
Bounds on Lorentz-violation parameters should be attainable in this
case at a level of $10^{-25}$ GeV.
However,
the analysis is still underway.

{\it Meson Experiments}:
In addition to these atomic experiments,
high-precision Lorentz and CPT tests have also been
performed for mesons in K, B, and D systems
\cite{mesonexpt,mesonth}.
In these systems the only relevant SME parameters are $a_\mu$.
Bounds on the order of $10^{-21}$ GeV have been obtained.

{\it Spin-Polarized Torsion Pendulum}:
A recent experiment at the University of Washington used a spin-polarized
torsion pendulum to achieve high sensitivity to
Lorentz violation in the electron sector
\cite{eotwash}.
Its sensitivity stems from the combined effect of a large number
of aligned electron spins.
The experiment uses stacked toroidal magnets that have a net
electron spin $S \simeq 8 \times 10^{22}$,
but which have a negligible magnetic field.
The pendulum is suspended on a turntable and a time-varying
harmonic signal is sought.
An analysis of this system reveals that in addition to a signal with the
period of the rotating turntable,
the effects of Lorentz and CPT violation induce additional
time variations with a sidereal period caused by Earth's rotation.
The group at the University of Washington has analyzed their data
and has obtained a bound on the electron parameters
equal to $|\tilde b_J^e| \lsim 10^{-29}$ GeV for $J=X,Y$ and
$|\tilde b_Z^e| \lsim 10^{-28}$ GeV
\cite{eotwash}.

\section{Conclusions}

Several new tests of Lorentz and CPT symmetry have been performed 
in recent years in a variety of particle systems.
As sharp as the bounds from these experiments are,
there is still room for improvement,
and it is likely that the next few years will continue to provide
increasingly sharp new tests of Lorentz and CPT symmetry in atomic systems.
In particular,
it should be possible to obtain bounds on many of the parameters
that current experiments have not probed.
One promising approach is to conduct atomic clock-comparison tests
in a space satellite
\cite{bklr02}.
These will have several advantages over traditional
ground-based experiments, 
which are typically insensitive to the
direction $Z$ of Earth's axis and ignore boost
effects associated with timelike directions.
For example,
a clock-comparison experiment conducted aboard 
the International Space Station (ISS)
would be in a laboratory frame that is both rotating and boosted.
It would therefore immediately gain sensitivity to
both the $Z$ and timelike directions.
This would more than triple the number of Lorentz-violation
parameters that are accessible in a clock-comparison experiment.
Since there are several missions already planned for
the ISS which will compare Cs and Rb atomic clocks and H masers,
the opportunity to perform these new Lorentz and CPT
tests is quite real.

In summary,
by using the general framework of the SME we are able to analyze 
Lorentz and CPT tests in a variety of atomic and particle experiments.
Many of the bounds that have been obtained are well within
the range of suppression factors associated with the Planck scale.

\noindent
{\bf ACKNOWLEDGMENTS}

I would like to acknowledge my collaborators
Alan Kosteleck\'y, Charles Lane, and Neil Russell.
This work was supported in part
by the National Science Foundation
under grant number PHY-0097982.

\bigskip




\end{document}